# Average probability of a dangerous failure on demand: Different modelling methods, similar results

**Florent Brissaud**[*], **Luiz Fernando Oliveira**
DNV France, Paris, France
*florent.brissaud@dnv.com

**Abstract:** According to the IEC 61508 functional safety standard, it is required to estimate the achieved safety integrity of the system due to random hardware failures. For a safety function operating in a low demand mode, this measure is the average probability of a dangerous failure on demand (PFDavg).
In the present paper, four techniques have been applied to various configurations of a case study: fault tree analyses supported by GRIF/Tree, multi-phase Markov models supported by GRIF/Markov, stochastic Petri nets with predicates supported by GRIF/Petri, and approximate equations (developed by DNV and different from those given in IEC 61508) supported by OrbitSIL.
It is shown that all these methods yield very similar results for PFDavg, taking the characteristics required by the standard into account. The choice of a method should therefore not be determined by dogmatic assumptions, but should result of a balance between modelling effort and objectives, given the system properties. For this task, a discussion about pros and cons of each method is proposed.

## 1. INTRODUCTION

To provide a generic approach for lifecycle activities regarding electrically-based systems used to perform safety functions, the IEC 61508 [1] functional safety standard has been introduced (the second edition has been available since 2010). Then, application sector standards were developed, such as the IEC 61511 [2] (currently under revision) for the process industry.

Once the system safety requirements are allocated and specified, in terms of safety function (SIF) and safety integrity (probability of satisfactorily performing the SIF), it is required (among other clauses) to estimate the achieved safety integrity of the system due to random hardware failures.

For a SIF operating in a low demand mode (only performed on demand, and no greater than one per year), the measure used is the average probability of a dangerous failure on demand ($PFD_{avg}$), computed as a mean unavailability. This measure has to take several characteristics into account: system architecture (e.g. $M$-out-of-$N$), failure rates (for detected and undetected failures), common cause failures (e.g. using beta-factor), intervals and effectiveness of tests (online diagnostic tests and periodic proof tests), and repair times. To this end, several methods are proposed, including fault tree analyses [3, 4, 5] (mathematically equivalent to reliability block diagrams), (multi-phase) Markov models [6, 7, 4, 5], (stochastic) Petri nets (with predicates) [3, 4, 5], and approximate equations [8, 9, 10, 11].

A case study is presented in Section 2, assuming different sets of parameters. In Section 3, fault tree analyses, Markov models, Petri nets, and approximate equations, are respectively applied to the various configurations of the case study. Results are reported in Section 4, with a discussion on the pros and cons of each method.

## 2. CASE STUDY

### 2.1. System description and assumptions

The system architecture is defined by a "$M$-out-of-$N$" configuration, that is, the system is composed of $N$ channels (i.e. subsystems) and is able to perform its function if any $M$ or more channels (of $N$) are in an operating mode. This system is actually a subpart (i.e. sensors, logic solver, or final elements) of a safety instrumented system (SIS) and to assess the entire SIF, it has to be completed by the remaining parts.

The function of the system is assumed to be a safety function that is only performed on demand, and the frequency of demands is no greater than one per year. This frequency of undesired demands is low enough to reasonably ignore the effects of these demands on the system availability (in practice, this frequency is commonly between 1 per 10 years to 1 per 10000 years).

Each channel that composes the system is assumed to be in an operating mode if and only if it is not in a dangerous failure mode. For each channel, three (dangerous) failure modes are considered:
▪ dangerous failure detected online (i.e. as soon as it occurs) by diagnostic tests;
▪ dangerous failure only detected by (periodic) proof tests, (and also detected by real demands);
▪ dangerous failure only detected by real demands (since the proof tests are not *100*% effective).
Each failure mode is assumed to occur according to a constant failure rate (i.e. exponential distribution), and the failure rates are the same for all channels (but, of course, depend on the failure mode). A factor of common cause failures (CCF) is also assumed for each failure mode (i.e. conditional probability that the failure mode occurs for all channels that are in operating mode, as soon as it occurs for one channel). All channels are assumed to be in an operating mode at time $t_0 = 0$.

As soon as a failure mode is detected (online by a diagnostic test, or periodically by a proof test), it is repaired according to a time to restoration defined by a constant repair rate (i.e. exponential distribution), and the repair rates are the same for all channels (but depend on the failure mode). It is assumed that the required resources for maintenance are always available, that is, no logistic delay or crew unavailability is considered outside the definition of the repair rates. When a channel is in a dangerous failure mode, it remains in this state up to the end of the repair and no compensating measures are assumed during repair.

In order to detect failures that are not detected online (by diagnostic tests), proof tests are performed periodically and for all channels at the same time (no staggered testing). However, these proof tests are not necessarily *100*% effective since a part of possible failures can only be detected by real demands. The latter failures are assumed never detected (no real demands) in the period of time that is considered. Testing duration is assumed to have a negligible impact on safety integrity (and it is therefore not quantified).

The safety integrity of the system is computed based on a period of time which is, for example, a period of overhaul testing (which can be performed by applying a real demand), or the planned system lifetime.

## 2.2. Notations

$MooN$  system architecture i.e. the system is composed of $N$ channels and is able to perform its function if any $M$ or more channels (of $N$) are in an operating mode

$\lambda_D$  failure rate of any channel (which composes the system), regarding dangerous failures
$\lambda_{DD}$  failure rate of any channel, regarding dangerous failures detected online by diagnostic tests
$\lambda_{DU}$  failure rate of any channel, regarding dangerous failures undetected online
$\lambda_{DUT}$  failure rate of any channel, regarding dangerous failures only detected by proof tests
$\lambda_{DUU}$  failure rate of any channel, regarding dangerous failures only detected by real demands

$DC$  diagnostic test coverage, such as $\lambda_{DD} = DC \times \lambda_D$ and $\lambda_{DU} = (1 - DC) \times \lambda_D$
$PTC$  proof test coverage, such as $\lambda_{DUT} = PTC \times \lambda_{DU}$ and $\lambda_{DUU} = (1 - PTC) \times \lambda_{DU}$

$\beta_{DD}$  factor of common cause failures, regarding dangerous failures detected online by diagnostic tests
$\beta_{DUT}$  factor of common cause failures, regarding dangerous failures only detected by proof tests
$\beta_{DUU}$  factor of common cause failures, regarding dangerous failures only detected by real demands

$\mu_{DD}$  repair rate of any channel, regarding dangerous failures detected online by diagnostic tests
$\mu_{DUT}$  repair rate of any channel, regarding dangerous failures only detected by proof tests

$T_1$  period of proof tests
$T_0$  period of computation for the safety integrity of the system

$PFD_{avg}$  average probability of a dangerous failure on demand (computed on period $T_0$)

Each failure rate can therefore be divided in two parts: independent failures (no related to common cause failures), and (due to) common cause failures (CCF). These parts are respectively:

- *(1 – $\beta_{DD}$) × $\lambda_{DD}$* and *$\beta_{DD}$ × $\lambda_{DD}$* for dangerous failures detected online by diagnostic tests;
- *(1 – $\beta_{DUT}$) × $\lambda_{DUT}$* and *$\beta_{DUT}$ × $\lambda_{DUT}$* for dangerous failures only detected by proof tests;
- *(1 – $\beta_{DUU}$) × $\lambda_{DUU}$* and *$\beta_{DUU}$ × $\lambda_{DUU}$* for dangerous failures only detected by real demands.

## 2.3. Sets of parameters

To compare various configurations of a case study, the six sets of parameters given in Table 1 are assumed.

|  | **case *i*** | **case *ii*** | **case *iii*** | **case *iv*** | **case *v*** | **case *vi*** |
|---|---|---|---|---|---|---|
| **MooN** | 1oo1 | | 1oo2 | | 2oo3 | |
| **$\lambda_D$ [hour$^{-1}$]** | $2.70 \times 10^{-6}$ | $1.35 \times 10^{-5}$ | $2.70 \times 10^{-6}$ | $1.35 \times 10^{-5}$ | $2.70 \times 10^{-6}$ | $1.35 \times 10^{-5}$ |
| **DC** | 0.50 | 0.25 | 0.50 | 0.25 | 0.50 | 0.25 |
| **PTC** | 0.90 | 0.70 | 0.90 | 0.70 | 0.90 | 0.70 |
| **$\beta_{DD}$** | 0.02 | 0.05 | 0.02 | 0.05 | 0.02 | 0.05 |
| **$\beta_{DUT}$** | 0.05 | 0.10 | 0.05 | 0.10 | 0.05 | 0.10 |
| **$\beta_{DUU}$** | 0.05 | 0.10 | 0.05 | 0.10 | 0.05 | 0.10 |
| **$\mu_{DD}$ [hour$^{-1}$]** | 0.0417 | 0.0833 | 0.0417 | 0.0833 | 0.0417 | 0.0833 |
| **$\mu_{DUT}$ [hour$^{-1}$]** | 0.0417 | 0.0833 | 0.0417 | 0.0833 | 0.0417 | 0.0833 |
| **$T_1$ [hours]** | 4,383 | 8,766 | 4,383 | 8,766 | 4,383 | 8,766 |
| **$T_0$ [hours]** | 70,128 (i.e. 8 years) | | | | | |

Table 1. Sets or parameters for the case study

## 3. METHODS APPLICATION

### 3.1. Fault tree analyses supported by GRIF/Tree (ALBISIA)

Fault trees are deductive techniques which express top events as combinations of basic events, using logic gates such as "or", "and", and "*MooN*." Analyses are performed using Boolean algebra, usually through the minimal cut sets (MCS), (i.e. minimal sets of basic events whose occurrence ensure the top event occurrence). Among common warnings for fault tree analyses it should be noted that: combining the average time-dependent probabilities of basic events does not provide the average time-dependent probability of top event (the top event probability has to be computed according to time, and the average has to be computed afterwards); if cut-offs of MCS are used (to save time when analysing large systems), the relevant cut sets may be time-dependent [3]. To perform fault tree analyses, the software tool is therefore important. In the present paper, the Tree module of GRIF [12] is used. It is based on ALBIZIA, a Binary Decision Diagram (BDD) computation engine developed by Total.

An intrinsic feature of fault trees is that all the basic events are independent (except, of course, for repeated basic events). It means that if a basic event occurs, it cannot prevent the occurrence of another basic event. For example, a failure mode of a channel does not prevent the other failure modes of the same channel (i.e. multiple states at the same time are possible), and an independent failure does not prevent a common cause failure. This could yield to an overestimation of the top event probabilities if the basic events would be assumed as not compatible, and especially if the latter are numerous (e.g. numerous failure modes). In addition, (basic) fault trees are "static models," that is, the model architecture (e.g. logic gates) does not depend on time and/or stochastic variables. However, to model top events that depend on the order of occurrence (i.e. sequences) of the basic events, dynamic fault trees have been also developed [13].

The fault tree applied to the case study with *1oo2* architecture (cases *iii* and *iv*, cf. Table 1) is provided in Figure 1. The top event represents the failure of the safety function at the system level. Then, several gates depict the possible failures of the channels that compose the system. At the lowest level, the basic events represent the independent (one for each channel) and common cause (one for all channels) failures, regarding dangerous failures detected online by diagnostic tests, only detected by proof tests, and only detected by real demands (i.e. nine basic events in total).

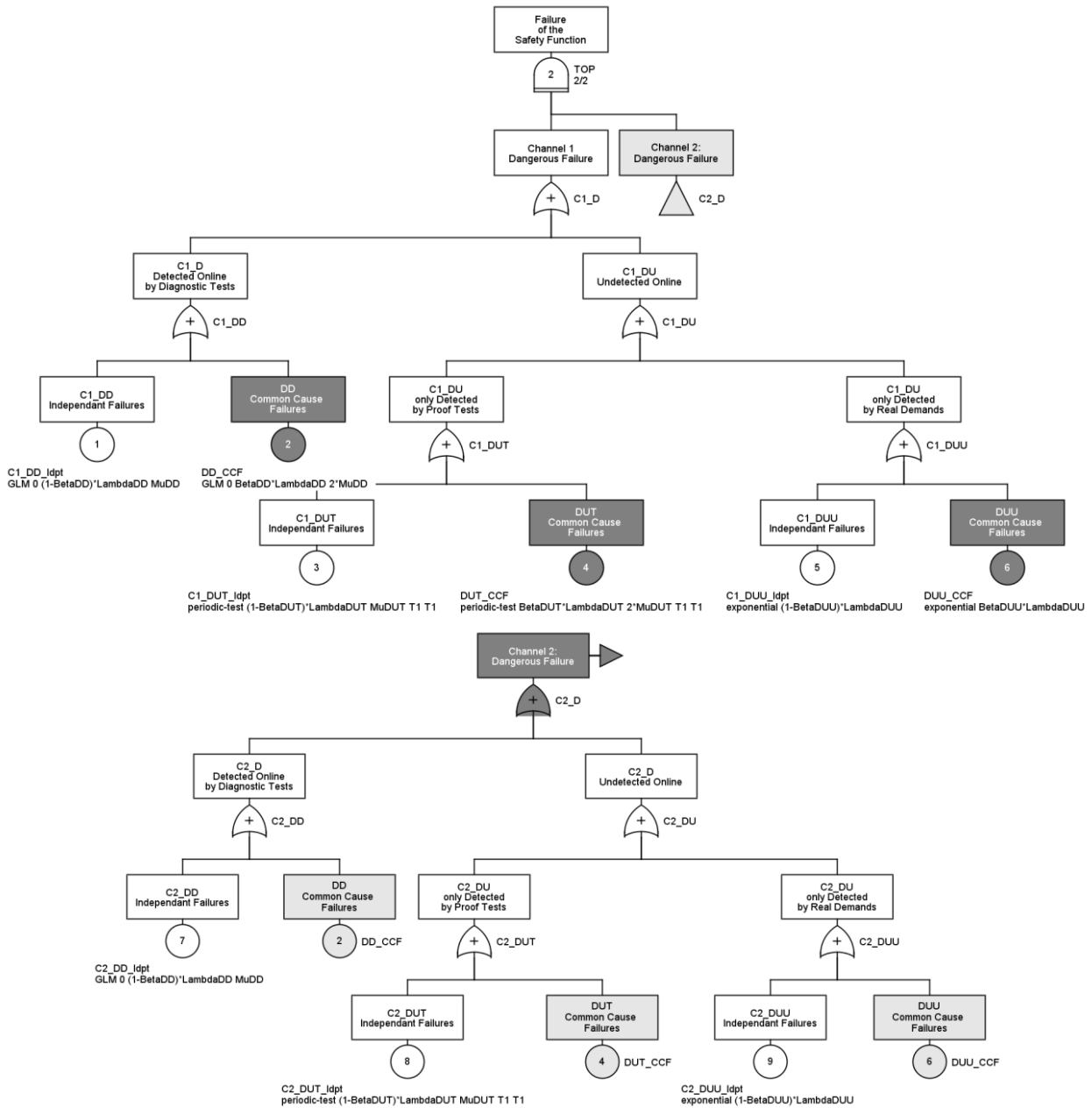

Figure 1. Fault tree with GRIF/Tree, case study with *1oo2* architecture

Using GRIF/Tree, a gate with a triangle on the right means a transfer to the corresponding gate with a triangle at bottom (e.g. Gates C2_D, in Figure 1). Moreover, a basic event that is light grey filled is a repetition of the corresponding basic event that is dark grey filled (e.g. Basic Events 2, 4, and 6, in Figure 1). A basic event and its repetition are physically the same events. Finally, the top gate depicted in Figure 1 is a "*MooN*" gate. In this figure, "2/2" means "*2oo2*" (i.e. equivalent to an "and" gate) – since a fault tree is failure-oriented and not success-oriented, this gate is rightly used to model a *1oo2* architecture. For the basic events, three kinds of laws have been used:
▪ "GLM," which corresponds to failures that are detected on-line, defined by a probability of failure to start (not used here, so it is equal to *0*), a failure rate, and a repair rate;
▪ "periodic-test," which corresponds to failures that are periodically detected, defined by a failure rate, a repair rate, the time interval between two consecutive tests, and the date of the first test (here equal to the previous parameter, otherwise staggered tests can be assumed);
▪ "exponential," which corresponds to failures that are never detected, defined by a failure rate.
Other laws are available in GRIF/Tree, which allows modelling additional features such as Weibull laws, imperfect tests/repairs, test/repair-generated failures, unavailability/degraded modes during testing, etc.

Note that the repair rates for common cause failures are equal to the number of channels multiplied by the repair rate of one channel (since it is assumed that the required resources for maintenance are always available, cf. Section 2.1). In fact, since all the basic events are independent, when at least one channel is repaired, the basic event that represents the common failure of all channels is not valid anymore.

## 3.2. Multi-phase Markov models supported by GRIF/Markov (MARK-XPR)

Markov models are state-based approaches that allow mathematical analyses, such as the computation of time-dependent probabilities of being in given states, average frequencies of entering in given states, and average sojourn times.

A Markov graph is made up of states (circles) and transitions (arrows). The states are defined at the system level. For multi-component systems, the states are therefore defined by the possible state combinations of system components. The number of states in a Markov model therefore depends on the number of components and on the number of operating/failure/repair/… modes that has to be assumed for each component. To reduce the combinatorial problem, it is convenient to assume that each component cannot be in multiple modes at the same time (e.g. a failure mode of a component prevents the other failure modes of the same component, contrarily to fault trees, cf. Section 3.1). Moreover, it is also often suitable to group together several combinations in one state. For example, if all system components are identical (e.g. same failure and repair rates), it may be possible to reason in terms of total number of each mode among all the components, instead of specific mode for each component. At time $t_0 = 0$, the initial state is defined by probabilities (e.g. the state where all system components are operating is usually the initial state with a probability of *1*). The transitions between states (e.g. failures, repairs) are then modelled by transition rates, which have to be constant (i.e. exponential distributions). The resulting Markov assumption is that the state at time $t+\Delta t$ depends on the state at time $t$, but not on time $t$, and not on the states before time $t$.

Because periodic tests occur at deterministic time instants (and therefore do not fit exponential distributions), multi-phase Markov models have been developed. The system is then modelled through different phases (e.g. operation, testing, repair, etc.), using one Markov graph for each phase. The sequence of the phases is determined (as a cycle), as well as the constant duration of each phase. Moreover, when moving from one phase to the next one, a (probabilistic) linking matrix is used to define the next initial state (in the next phase) according to the current state (in the current phase). In the present paper, the Markov module of GRIF [12] is used. It is based on MARK-XPR, a multi-phase Markov computation engine developed by Total.

The multi-phase Markov model applied to the case study with *1*oo*2* architecture (cases *iii* and *iv*, cf. Table 1) is provided in Figure 2. State 1 is the initial state (depicted in colour), where both system channels are in an operating mode (denoted "OK"). The other states are defined by the possible combinations of failure modes (denoted DD, DUT, and DUU, following the notations described in Section 2.2) and repair modes (denoted RepDUT for the DUT repair modes), (the DD repair modes are the same as the DD failure modes since these failures are detected online, and the DUT repair modes are not relevant given the assumptions given in Section 2.1). Since the channels are identical (i.e. same modes and same rates), the system states are defined independently of the channel that causes the operating/failure/repair mode (e.g. "OK_DD" correspond to one channel in an operating mode and one channel in a DD failure mode, independently of which channel for which mode). Using GRIF/Markov, the "Eff." parameters attached to each state is the property that is computed. It this case, this is the probability of a dangerous failure on demand, which is equal to *0* if at least one channel is in an operating mode, and *1* otherwise (according to the *1*oo*2* architecture).

For this case study, only one phase is required. Its duration is equal to the period of proof tests. Once this period is done, the phase restarts. Moreover, at each restart, the linking matrix is used to transfer any failure mode detected by proof test (at the end of the phase), to the corresponding repair mode (at the beginning of the next phase). This linking matrix only uses probabilities equal to *0* or *1*. However, the use of linking matrices with probabilities between *0* and *1* and/or more than one phase allow modelling additional features such as staggered tests, imperfect tests/repairs, test/repair-generated failures, unavailability/degraded modes during testing, etc. The size of the model can however increase drastically when additional features are included; making the model, in some cases, unmanageable.

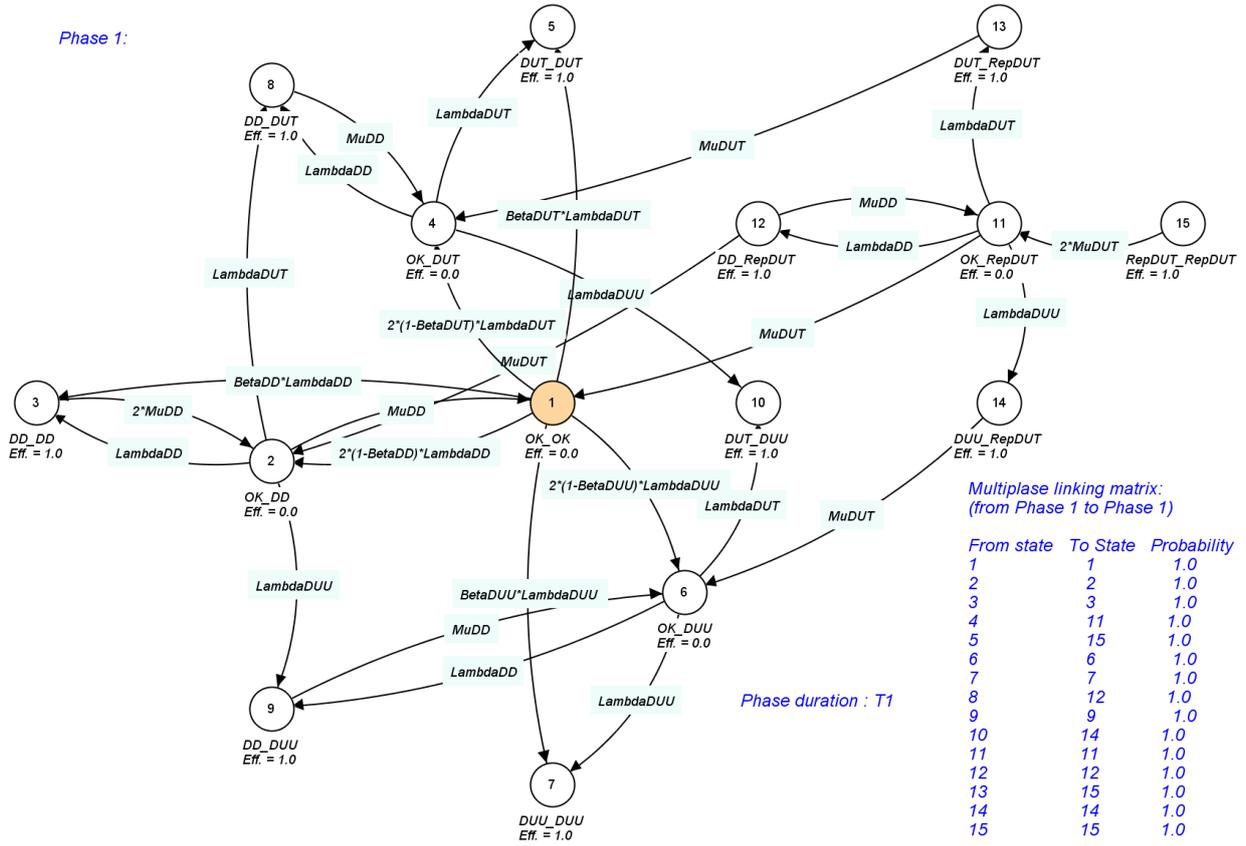

Figure 2. Markov model with GRIF/Markov, case study with *1*oo*2* architecture

### 3.3. Stochastic Petri nets with predicates supported by GRIF/Petri (MOCA-RP V13)

Petri nets provide a graphical tool for the behavioural modelling of (dynamical) systems, and then availability analyses can be performed by Monte Carlo simulations (i.e. results are obtained statistically from several simulated histories).

A basic Petri net is made up of places (circles) and transitions (rectangles). Connections (directed arcs) may link a place to a transition (input arc) or vice-versa (output arc), and may be "valued" (otherwise the value is assumed to be one). Places may contain tokens (small filled circles) which are "moved" through the enabled transitions when the latter are fired. A transition is enabled when each of its input places (linked to the transition by an input arc) contains a number of tokens equal to or greater than the corresponding input arc value. Firing an enabled transition then consists in two steps: first, removing, in each input place, a number of tokens equal to the corresponding input arc value; second, depositing, in each output place, a number of tokens equal to the corresponding output arc value.

Usually, the places of a Petri net represent objects or conditions, the tokens specify the values of these objects or conditions, and the transitions model the system activities. The time-dimension is introduced by time delays for firing transitions (given that the transitions remain enabled during these delays). In stochastic Petri nets, these delays are random variables. In addition, Petri nets with predicates use variables to include two other properties: guards, which are Boolean variables/expressions that disable transitions when not verified; and affectations, which are assignments that modify values of variables when transitions are fired. In the present paper, the Petri module of GRIF [12] is used. It is based on MOCA-RP, a high-speed Monte Carlo simulation engine developed by Total.

The Petri net applied to the case study with *1*oo*2* architecture (cases *iii* and *iv*, cf. Table 1) is provided in Figure 3 (with the tokens as defined at time $t_0 = 0$). Places 1 to 5 modelled the first system channel, Places 12 to 16 the second channels, and Places 6 to 11 the common cause failures.

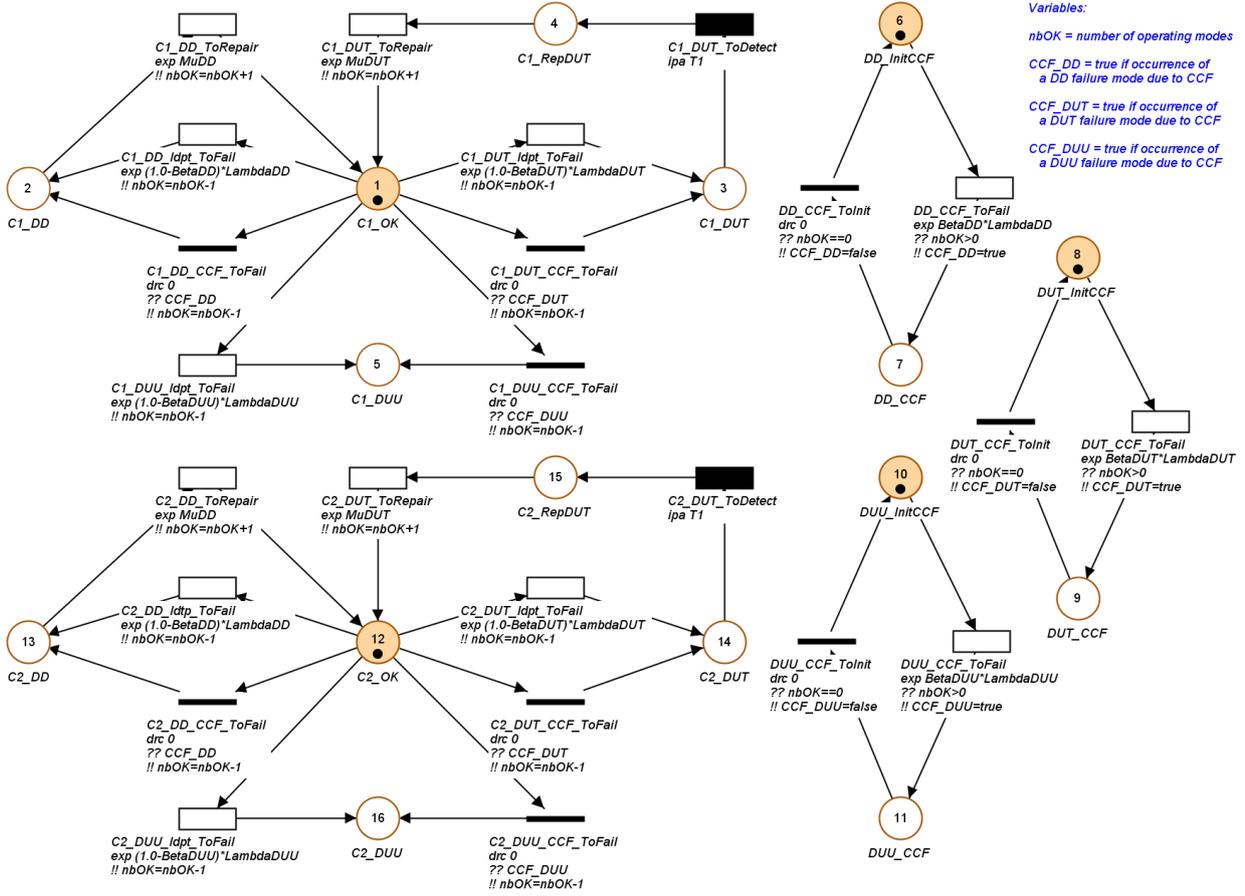

Figure 3. Petri net with GRIF/Petri, case study with *1oo2* architecture

Using GRIF/Petri, the first row below a place or a transition is the description of the element. For example, when a place with the description "Cx_y" (for Places 1 to 5 and 12 to 16) contains a token, it means that "channel x is in the operating/failure/repair mode y" (following the notations described in Section 2.2). (Here, it has been assumed that each channel cannot be in multiple modes at the same time, similar to Markov models, cf. Section 3.2.) A place that contains at least one token is depicted in colour (for this case study, a place can contain only *0* or *1* token). Below a transition, the expression starting with "??" is a guard, and the expression starting with "!!" is an affectation. For example, the transition from Place 6 to Place 7 can be fired only if the variable "nbOK" is greater than *0* (i.e. at least one channel is in an operating mode) and, when it is fired, the variable "CCF_DD" takes the value "true" (i.e. occurrence of a DD failure mode due to CCF). Since "CCF_DD=true," the transition from Place 1 to Place 2 (respectively, from Place 12 to Place 13) is therefore fired if Place 1 (respectively, Place 12) contains a token. If this transition is fired, the variable "nbOK" is reduced by *1* (i.e. the number of channel in an operating mode is reduced by *1*). If "nbOK" becomes equal to *0* (i.e. no channel is in an operating mode), then the transition from Place 7 to Place 6 resets the variable "CCF_DD" to "false." Using GRIF/Petri, a transition with a time delay that is exponentially distributed is depicted in white, a transition with no time delay is depicted by a flat black rectangle, and other deterministic transitions are depicted by a large black rectangle. For the transition delay, three kinds of laws have been used (given in the second row below the transitions):

▪ "exp," which is the Exponential distribution, defined by a rate (e.g. failure or repair rate);
▪ "drc," which is the Dirac distribution (i.e. deterministic time delay), defined by a constant time delay (here equal to *0* in order to have instantaneous transition if guards are fulfilled);
▪ "ipa," which is the "in advance schedule times," defined by a period at which the transition is enabled (used with the period of proof tests in order to be able to detect DUT failure modes at deterministic time instants).

Other laws are available in GRIF/Tree, including Uniform, Triangular, Weibull, Normal, and Log-Normal distributions. In addition, and because these Petri nets are behavioural modelling combined with simulation-based analyses, they provide a way to model (any?) additional features, including staggered tests, imperfect tests/repairs, test/repair-generated failures, unavailability/degraded modes during testing, etc.

### 3.4. Approximate equations supported by OrbitSIL

Approximate equations are based on Taylor series (usually of first order) of the exponential functions to approach reliability equations by formulas that contain only basic operators (e.g. no integral) in order to be computed by simple calculators. Since the Taylor series of first order overestimates exponential unreliability functions, these approximate equations are usually assumed as "conservative" (e.g. overestimation of $PFD_{avg}$). However, it should be noted that these equations are developed under specific assumptions and applying them outside the predefined scope can result to (dangerous) inaccurate results (e.g. when equations only assumed proof tests that are *100*% effective).

The above formulas have been developed by DNV and are different from those given in the IEC 61508 standard (in Part 6) [1]. These formulas have been proposed by L.F. Oliveira *et al.* (notably with the inclusion of partial proof tests [8] and common cause failures [9]) to be conceptually more consistent than those of IEC 61508 [9]. However, these formulas are also more "complicated" and the (non-commercial) software tool OrbitSIL has therefore been developed to compute them. Because it is based on Taylor series of first order, the following approximate equation that is used to compute $PFD_{avg}$ is considered as "valid" only if $\lambda_{DUT} \times T_1 < 0.1$ and $\lambda_{DUU} \times T_0 < 0.1$ (cf. the assumptions and notations given in Section 2):

$$PFD_{avg} = \binom{N}{N-M+1} \times \left((1-\beta_{DD}) \times \frac{\lambda_{DD}}{\mu_{DD}}\right)^{N-M+1} \quad (1)$$

$$+ \binom{N}{N-M+1} \times \left((1-\beta_{DUT}) \times \lambda_{DUT}\right)^{N-M+1} \times T_1^{N-M} \times \left(\frac{T_1}{N-M+2} + \frac{1}{\mu_{DUT}}\right) + \binom{N}{N-M+1} \times \left((1-\beta_{DUU}) \times \lambda_{DUU}\right)^{N-M+1} \times T_0^{N-M} \times \left(\frac{T_0}{N-M+2}\right)$$

$$+ \sum_{i=1}^{N-M} \binom{N-i}{N-M+1-i} \times \left(f(N-M+1-i,\beta_{DD}) \times \frac{\lambda_{DD}}{\mu_{DD}}\right)^{N-M+1-i} \times \binom{N}{i} \times (f(i,\beta_{DUT}) \times \lambda_{DUT})^i \times T_1^{i-1} \times \left(\frac{T_1}{i+1} + \frac{1}{\mu_{DUT}}\right)$$

$$+ \sum_{i=1}^{N-M} \binom{N-i}{N-M+1-i} \times \left(f(N-M+1-i,\beta_{DD}) \times \frac{\lambda_{DD}}{\mu_{DD}}\right)^{N-M+1-i} \times \binom{N}{i} \times (f(i,\beta_{DUU}) \times \lambda_{DUU})^i \times T_0^{i-1} \times \left(\frac{T_0}{i+1}\right)$$

$$+ \sum_{i=1}^{N-M} \binom{N-i}{N-M+1-i} \times \left((1-\beta_{DUT}) \times \lambda_{DUT}\right)^{N-M+1-i} \times T_1^{N-M-i} \times \left(\frac{T_1}{N-M+2-i} + \frac{1}{\mu_{DUT}}\right) \times \binom{N}{i} \times \left((1-\beta_{DUU}) \times \lambda_{DUU}\right)^i \times T_0^{i-1} \times \left(\frac{T_0}{i+1}\right)$$

$$+ \sum_{i=1}^{N-M-1} \sum_{j=1}^{N-M-i} \binom{N-i-j}{N-M+1-i-j} \times \left(f(N-M+1-i-j,\beta_{DD}) \times \frac{\lambda_{DD}}{\mu_{DD}}\right)^{N-M+1-i-j}$$

$$\times \binom{N-i}{j} \times \left((1-\beta_{DUT}) \times \lambda_{DUT}\right)^j \times T_1^{j-1} \times \left(\frac{T_1}{j+1} + \frac{1}{\mu_{DUT}}\right) \times \binom{N}{i} \times \left((1-\beta_{DUU}) \times \lambda_{DUU}\right)^i \times T_0^{i-1} \times \left(\frac{T_0}{i+1}\right)$$

$$+ \beta_{DD} \times \frac{\lambda_{DD}}{\mu_{DD}} + \beta_{DUT} \times \lambda_{DUT} \times \left(\frac{T_1}{2} + \frac{1}{\mu_{DUT}}\right) + \beta_{DUU} \times \lambda_{DUU} \times \left(\frac{T_0}{2}\right)$$

with the function $f(x,b)$ defined by:

$f(x,b) = \begin{cases} 1, & \text{if } x = 1 \\ 1 - b, & \text{if } x \neq 1 \end{cases}$

and the following combinatorial function:

$\binom{n}{k} = \frac{n!}{k! \times (n-k)!}$

In accordance with the assumptions given in Section 2.1, approximate Equation (1) does not consider any repair time regarding dangerous failures only detected by real demands, but it can be easily extended to include this parameter. However, additional features such as heterogeneous failure and repair rates, staggered tests, and test/repair-generated failures are more difficult to consider in such equations. Under other assumptions, and notably when the times to repair do not need to be modelled (e.g. when compensating measures are assumed during repair to maintain or restore the safety function), other approximate and "exact" equations have been developed, which can also include non-periodic tests [11].

### 4. DISCUSSIONS

#### 4.1. Results

The results obtained by each method, for the various configurations of the case study presented in Section 2, are reported in Table 2. Since the Markov graphs would contain 35 states, the authors have been discouraged from applying Markov models to cases *v* and *vi*. For Petri nets, $10^8$ simulations have been used for each case, resulting *90%* confidence intervals from *±0.03%* (for case *ii*) to *±0.5%* (for case *iii*). The computation time was between 40 minutes (for cases *i* and *ii*) and 75 minutes (for cases *v* and *vi*) with a 2.67 GHz processor, 4.00 GB of RAM.

|  | **case *i*** | **case *ii*** | **case *iii*** | **case *iv*** | **case *v*** | **case *vi*** |
|---|---|---|---|---|---|---|
| **Fault trees** | $7.43 \times 10^{-3}$ | $1.27 \times 10^{-1}$ | $4.31 \times 10^{-4}$ | $2.93 \times 10^{-2}$ | $5.48 \times 10^{-4}$ | $5.59 \times 10^{-2}$ |
| **Markov** | $7.41 \times 10^{-3}$ | $1.24 \times 10^{-1}$ | $4.29 \times 10^{-4}$ | $2.83 \times 10^{-2}$ | - | - |
| **Petri nets** | $7.41 \times 10^{-3}$ | $1.24 \times 10^{-1}$ | $4.30 \times 10^{-4}$ | $2.83 \times 10^{-2}$ | $5.47 \times 10^{-4}$ | $5.43 \times 10^{-2}$ |
| **Equations** | $7.46 \times 10^{-3}$ | $1.38 \times 10^{-1}$ | $4.31 \times 10^{-4}$ | $3.25 \times 10^{-2}$ | $5.49 \times 10^{-4}$ | $6.98 \times 10^{-2}$ |

Table 2. Results: $PFD_{avg}$, for cases *i* to *vi*

Multi-phase Markov models and stochastic Petri nets with predicates produce the same results (the difference for case *iii* is not significant according to the confidence interval) because the same assumptions have been done. Results obtained by fault tree analyses are slightly greater (maximum of *3.5%* for case *iv*) due to the independency of basic events, which allows multiple failure modes of the same channel at the same time (and these combinations was excluded in Markov models and Petri nets). Finally, approximate equations are the more overestimating (maximum of *28.5%* for case *vi*, compared to Petri nets) due to the mathematical approximations that have been used to get "simple" formulas (note that for cases *ii*, *iv*, and *vi*, $\lambda_{DUU} \times T_0 = 0.213$, which is greater than *0.1* and does not comply with the defined condition of validity).

To sum up, different methods which are fundamentally different (from a practicable and mathematical point of view) are able to provide very similar results for the average probability of a dangerous failure on demand ($PFD_{avg}$), taking the characteristics required by the IEC 61508 standard into account. To this end, it is however required to know the method that is used well (including its intrinsic assumptions and modelling capabilities) and to use an appropriate and efficient software tool (which is rightly based on the reliability theories). The choice of a method should therefore not be determined by dogmatic assumptions, but should result of a balance between modelling effort and objectives, given the properties of the system to be modelled. For this task, a discussion about pros and cons of each method is proposed in the next subsection.

## 4.2. Pros and cons of each method

|  | **Fault tree analyses** | **Multi-phase Markov models** | **Stochastic Petri nets with predicates** | **Approximate Equations** |
|---|---|---|---|---|
| **Model** | | | | |
| size of the model (for large systems) | linearly dependent | exponentially dependent | linearly dependent | fixed |
| time to model (for many systems) | relatively fast | quite long | quite long but can be reduced by prototypes | fast (one formula to be repeated) |
| less prone to modelling errors | simple models / limited possible errors | "rich" models / several possible errors | "rich" models / several possible errors | basic model / only few possible errors |
| model readability | easy for engineers | reserved to specialists | reserved to specialists | "black box" difficult to read |
| **Analysis** | | | | |
| analysis method | exact Boolean approach | exact mathematical approach | Monte Carlo simulations | approximate approach |
| time to analyse / uncertainty analyses | very fast with good algorithms | very fast with good algorithms | longer due to simulations | very fast |
| adapt to importance factor analyses | several dedicated importance factors | to be adapted | to be adapted | not adapted |
| **Flexibility** | | | | |
| heterogeneous failure/repair rates | good ability | ability but increase drastically the model | good ability | not adapted |
| advanced test/repair properties | ability with powerful software | ability but increase drastically the model | good ability | not adapted |
| any distributions (e.g. Weibull, Normal) | ability with powerful software | not adapted | good ability | not adapted |
| dynamic features (e.g. event sequences) | not adapted | ability for basic cases | good ability | not adapted |

Table 3. Comparison of the methods

Table 3 provides some criteria (without pretending to be exhaustive) to compare fault tree analyses, multi-phase Markov models, stochastic Petri nets with predicates, and approximate equations, as presented in Section 3 (note that these methods have also other extends that are not discussed in the present paper). These ratings are based on the experience of the authors, but remain subject to discussion.

To conclude:
▪ Approximate equations provide a fast and simple way to assess many simple/basic systems. However, this approach is also, by nature, the least flexible and it may be hazardous to use these equations outside the predefined assumptions. A drawback is therefore the potential for some users to apply such attractable formulas without the required cautions (especially in face of "non-conservative" assumptions).
▪ Fault tree analyses carry most of the advantages for engineers: easy to apply and to read, based on efficient analysis approaches, and allow modelling a lot of features (including advanced test and repair properties) – given that an efficient software tool is used. The only limitation concerns "dynamic features" (e.g. sequences of events) that cannot be modelled due to intrinsic assumptions (e.g. the independency of basic events).
▪ Stochastic Petri nets with predicates provide the most flexible approach to model any complex systems, notably those with specific features such as dynamic properties. This flexibility is allowed by the simulation-based analyses. However, this computational approach is also a drawback due to the required time to perform the analyses. Petri nets should therefore be recommended for cases that cannot be treated properly by fault trees.
▪ Multi-phase Markov models have no real advantage compared to other methods, and have probably the lowest cost/benefit ratio in terms of modelling effort versus modelling abilities.